\documentclass[notitlepage,a4paper,aps,prd,onecolumn,superscriptaddress,nofootinbib,groupedaddress]{revtex4-1}
\usepackage{enumerate}
\usepackage{dsfont}
\usepackage{amsfonts}
\usepackage{amssymb}
\usepackage[latin1,utf8]{inputenc}
\usepackage[T1]{fontenc}
\usepackage{mathtools}
\usepackage{bbold}
\usepackage{ulem}

\usepackage[dvipsnames]{xcolor}

\usepackage[colorlinks=true]{hyperref}
\hypersetup{pdfstartview=FitH}

\newcommand{\D}[0]{\mathrm{D}}
\newcommand{\dd}[0]{\mathrm{d}}

\begin{document}
\title{Teleparallel theories of gravity as analogue of non-linear electrodynamics}

\author{Manuel Hohmann}
\email{manuel.hohmann@ut.ee}
\affiliation{Laboratory of Theoretical Physics, Institute of Physics, University of Tartu, W. Ostwaldi 1, 50411 Tartu, Estonia.}

\author{Laur J\"arv}
\email{laur.jarv@ut.ee}
\affiliation{Laboratory of Theoretical Physics, Institute of Physics, University of Tartu, W. Ostwaldi 1, 50411 Tartu, Estonia.}

\author{Martin Kr\v{s}\v{s}\'ak}
\email{martin.krssak@ut.ee}
\affiliation{Laboratory of Theoretical Physics, Institute of Physics, University of Tartu, W. Ostwaldi 1, 50411 Tartu, Estonia.}

\author{Christian Pfeifer}
\email{christian.pfeifer@ut.ee}
\affiliation{Laboratory of Theoretical Physics, Institute of Physics, University of Tartu, W. Ostwaldi 1, 50411 Tartu, Estonia.}

\begin{abstract}
The teleparallel formulation of gravity theories reveals close structural analogies to electrodynamics, which are more hidden in their usual formulation in terms of the curvature of spacetime. We show how every locally Lorentz invariant teleparallel theory of gravity with second order field equations can be understood as built from a gravitational field strength and excitation tensor which are related to each other by a constitutive relation, analogous to the premetric construction of theories of electrodynamics. We demonstrate how the previously studied models of $f(\mathbb{T})$ and $f(T_\text{ax},T_\text{ten},T_\text{vec})$ gravity as well as teleparallel dark energy can be formulated in this language. The advantage of this approach to gravity is that the field equations for different models all take the same compact form and general results can be obtained. An important new such result we find is a constraint which relates the field equations of the tetrad and the spin connection.
\end{abstract}

\maketitle

%%%%%%%%%%%%%%%%%%%%%%%%%%%%%%%%%%%%%%%%%%%%%%%%%%%%%%%
\section{Introduction}
Electrodynamics is described by the well-known Maxwell equations. In the language of differential forms these can be formulated in terms of the field strength $2$-form, a conserved current $3$-form, an excitation $2$-form as well as a relation between field strength and excitation called \textit{constitutive relation}. In the most simple case of Maxwell electrodynamics in vacuum this constitutive relation is given by the Hodge dual. In the premetric approach to electrodynamics \cite{Hehl} the constitutive relation is generalized to an arbitrary dependence of the excitation on the field strength and background fields, while the Maxwell equations retain their form. This generalization allows for the description of the vast physical phenomenology of the electromagnetic interaction from linear and non-linear effects in media to various non-linear theories of electrodynamics like Born-Infeld in one unified language.

Gravity, on the contrary, is described by general relativity (GR) in a completely different way. The Einstein equations determine the metric of spacetime from the matter energy-momentum through the curvature of the Levi-Civita connection of the metric. It is possible to construct various extended/modified theories of Einstein's gravity, which typically lead to structurally different field equations. See \cite{Sotiriou:2008rp,DeFelice:2010aj,Capozziello:2011et,Nojiri:2017ncd} for reviews.

An alternative description of the gravitational field and its dynamics is given on the basis of so-called teleparallelism, which allows to identify certain analogies between the dynamics of gravity and electromagnetism. In teleparallelism the Riemannian geometry of general relativity is replaced by the curvature-free teleparallel geometry and the gravitational interaction is attributed to the torsion of spacetime \cite{Einstein1928,Sauer:2004hj,Moller1961,Hayashi:1967se,Cho:1975dh,Hayashi:1977jd,Hehl:1994ue,AP}. It is possible to reformulate the dynamics of gravity known from general relativity equivalently into the teleparallel formalism, which is known as teleparallel gravity or teleparallel equivalent of general relativity (TEGR). This provides another starting point to construct further modified and extended theories of gravity~\cite{Ferraro:2006jd,Ferraro:2008ey,Bengochea:2008gz,Linder:2010py,Cai:2015emx,Geng:2011aj,Maluf:2011kf,Bahamonde:2017wwk}.

An interesting aspect of teleparallel gravity is that it arises as a gauge theory of the translation group \cite{Cho:1975dh,Hayashi:1967se,Hayashi:1977jd,Blagojevic,AP}, allowing to incorporate gravity into the gauge paradigm together with other fundamental interactions. As a consequence, the field equations of teleparallel gravity take a very similar form to those of the standard Maxwell or Yang-Mills theory. The torsion tensor can be viewed as the gravitational analogue of the electromagnetic field strength, while the so-called superpotential plays the role of the gravitational excitation tensor. This similarity has recently been used by Itin, Hehl and Obukhov~\cite{Itin:2016nxk} to demonstrate how one can generalize teleparallel gravity by choosing a general local and linear gravitational constitutive relation between the superpotential and the torsion. This leads to the gravity model known as New General Relativity~\cite{Hayashi:1979qx}, which includes teleparallel gravity as a special case. The whole construction can be seen as the gravitational analogue of the so-called premetric generalization of Maxwell electrodynamics \cite{Hehl:2016glb,Hehl:2016yle,Rubilar:2007qm,Hehl:2002hr,Rubilar:2001gj, Itin2018}, which is part of the framework of premetric electrodynamics. Note that another approach to understand teleparallel gravity in the line of gravitoelectromagnetism can be found in the literature \cite{Spaniol:2008rt, Ming:2017tna}, while a constitutive relation motivated by continuum mechanics is explored in \cite{Boehmer:2014jsa}.

In this paper we extend the analogy between general theories of electrodynamics and teleparallel gravity, formulated covariantly in terms of constitutive relations, in two directions:

Firstly we show that it is possible to realize every locally Lorentz invariant teleparallel theory of gravity with second order field equations in the aforementioned framework. We find that the field equations for all these models have a common structure, which is similar to Maxwell's equations in premetric electrodynamics. The difference between the theories lies solely in the choice of different gravitational constitutive laws which supplement the field equations. In particular, we construct the gravitational constitutive relation for previously studied modified teleparallel models~\cite{Ferraro:2006jd,Ferraro:2008ey,Bengochea:2008gz,Linder:2010py,Cai:2015emx,Geng:2011aj,Maluf:2011kf,Bahamonde:2017wwk}. Moreover, we use this framework to outline more general classes of teleparallel theories of gravity inspired by Pleba\'nski electrodynamics. Thus it is possible to understand different theories of gravity in a similar way as different theories of electrodynamics.

Secondly we include in our construction a non-vanishing teleparallel spin connection into the gravitational action along the lines of the recently proposed Lorentz covariant formulation of teleparallel theories
\cite{Krssak:2015rqa,Krssak:2015lba,Krssak:2015oua,Krssak:2017nlv}. The matter field action, i.e.\ the coupling between gravity and all other matter fields, is unaffected by the introduction of the spin connection since we do not change the minimal coupling prescription: matter couples to the metric and its Levi-Civita connection, which can be expressed as functions of the tetrads alone. The reason to consider only minimally coupled fields is to only introduce changes in the gravitational dynamics but not in the dynamics of the matter fields on a fixed background geometry. In consequence we show that for an arbitrary generalized model the result of the variation with respect to the spin connection is identical to the antisymmetric part of the tetrad field equations. This provides a constraint which relates the spin connection and the tetrad to each other and leaves only the symmetric part of the tetrad field equations determining the dynamics of the theory.

The outline of the article is as follows. We begin by recalling the premetric formulation of electrodynamics in section \ref{SecED}. In section \ref{ssec:TPFieldS} we summarize the teleparallel equivalent formulation of general relativity in the language of differential forms. Then we formulate gravity in the analogue language of electrodynamics in section \ref{ssec:TPInd}, which is one of the central results of this article. Afterwards, in section \ref{sec:GravConst} we illustrate how within this scheme several models of modified and extended teleparallel theories of gravity can be formulated in terms of different gravitational constitutive laws. Finally we discuss our findings and give an outlook for future developments in section \ref{sec:disc}. A detailed derivation of the gravitational field equations is presented in appendix \ref{app:feqder}.

We use the following conventions in this article: Greek indices $\mu,\nu,\ldots$ label the Lorentz (tetrad) indices and are raised and lowered with the Minkowski metric $\eta_{\mu\nu} = \textrm{diag}(1,-1,-1,-1)$. Latin indices $a,b,c,\ldots$ denote spacetime indices. Differential $m$-forms $\Omega$ are expanded in local coordinates with a factor $\tfrac{1}{m!}$,
\begin{align*}
	\Omega = \frac{1}{m!}\Omega_{a_1a_2...a_m}\ dx^{a_1} \wedge dx^{a_2} \wedge ... \wedge dx^{a_m}\,.
\end{align*}
Tensors with Greek index describe the gravitational interaction, tensors without a Greek index denote objects used in electrodynamics.

%%%%%%%%%%%%%%%%%%%%%%%%%%%%%%%%%%%%%%%%%%%%%%%%%%%%%%%
\section{Premetric Electrodynamics}\label{SecED}
General electrodynamics can be described by the field equations
\begin{align}
	\mathrm{d} F = 0, \quad \mathrm{d} H = J, \label{eq:dF=0dH=J}
\end{align}
where $F$ is the electromagnetic field strength $2$-form, $H$ is the excitation $2$-form and $J$ the closed current $3$-form. In the case of Maxwell electrodynamics the excitation form is related to the field strength form using the Hodge dual map and the vacuum impedance $\lambda_0 = \sqrt{\frac{\epsilon_0}{\mu_0}}$
\begin{align}\label{MaxED}
H= \lambda_0 \star F\,.
\end{align}
For the following arguments we will choose units such that the vacuum impedance is normalized $\lambda_0 = 1$, in SI units its value is $\lambda_0 = \frac{1}{377 \Omega}$.

The basic idea behind this premetric approach to electrodynamics is that the field equations \eqref{eq:dF=0dH=J} define every gauge invariant theory of electrodynamics which realizes electric charge and magnetic flux conservation. The only additional ingredient needed to make the theory predictive is a constitutive relation which expresses the excitation in terms of the fields strength $H=H(F)$ \cite{Hehl}. This constitutive relation can be of any kind and does not necessarily involve a metric tensor, as it is used in Maxwell electrodynamics on curved spacetime \eqref{MaxED}.

The specification of the constitutive relation $H(F)$ distinguishes different theories. The most familiar constitutive law is a local and linear relation $H=\kappa(F)$, which in local coordinates takes the form
\begin{align}\label{eq:edynloclin}
H_{ab} = \frac{1}{2}\kappa_{ab}{}^{cd}F_{cd} = \frac{1}{4}\epsilon_{abcd}\chi^{cdef}F_{ef}.
\end{align}
The map $\kappa$ maps $2$-forms to $2$-forms and is called the constitutive map while the object $\chi^{cdef}$ is the constitutive density and $\epsilon$ the totally antisymmetric Levi-Civita symbol. The fields $\kappa$ or $\chi$ equivalently define the class of theories called local and linear premetric electrodynamics.

As mentioned above Maxwell electrodynamics on a spacetime $(M,g)$ is defined by choosing $\kappa = \star$, the Hodge dual operator of the metric, which implies
\begin{align}\label{eq:maxconlaw}
\chi^{abcd} = 2 |g|^\frac{1}{2} g^{a[c}g^{d]b},
\end{align}
with $|g| = |\det g_{ab}|$.
Electrodynamics in media, like in an uniaxial birefringent crystal for example~\cite{Perlick,Pfeifer:2016har}, is described by the constitutive density
\begin{align}\label{eq:uniax}
\chi^{abcd} =|g|^\frac{1}{2} \big( 2 g^{a[c}g^{d]b} + 4 X^{[a}U^{b]} X^{[d}U^{c]}\big)\,,
\end{align}
where $X$ is the spacelike direction of the optical axis of the crystal and $U$ the timelike direction which characterizes the rest frame of the crystal.
Recently local and linear premetric electrodynamics has gained some interest as model for a quantum field theory based on a more general causal structure than that of a Lorentzian metric \cite{Rivera:2011rx,Pfeifer:2016har,Grosse-Holz:2017rdt,Fewster:2017mtt}.

In general there is no a priori need to choose the constitutive relation $\kappa$ to be linear. It is well-known that electrodynamics in media exhibits non-linear effects. The important and interesting point to make here is, that in order to capture and describe these effects it is not necessary to change the fundamental field equations \eqref{eq:dF=0dH=J}, but it suffices to change the constitutive relation $H(F)$, for example to a local and non-linear one. A famous class of non-linear electrodynamics are so-called Pleba\'nski theories~\cite{Hehl}. They are defined by a constitutive relation involving two general functions $U$ and $V$
\begin{align}\label{CRPleba\'nski}
H(F) = U(I_1, I_2) F + V(I_1, I_2) \star F\,,
\end{align}
where $I_1$ and $I_2$ are the scalars
\begin{align}
I_1 = \frac{1}{2}\star(F\wedge \star F),\quad I_2 = \frac{1}{2}\star(F\wedge F)\,.
\end{align}
The $\star$ denotes again the Hodge dual operator on a $4$-dimensional spacetime $(M,g)$.
Special instances of Pleba\'nski electrodynamics are the Born-Infeld theory \cite{Born425},
\begin{align}
	H(F) =  \frac{\star F+\frac{1}{f^2} I_2 F}{\sqrt{1 - \frac{1}{f^2}I_1 - \frac{1}{ f^4}I_{2}^2}}\,,
\end{align}
and the Heisenberg-Euler theory \cite{Heisenberg1936} emerging as counter term in quantum electrodynamics,
\begin{align}
H(F) =  \bigg[\bigg(1 + \frac{16 \alpha}{45 B^2} I_1\bigg)\star F +\frac{28 \alpha}{45 B^2} I_2F\bigg]\,.
\end{align}
At the end of this article we will find the teleparallel theories of gravity which are analogs to the Pleba\'nski theories of electrodynamics.

%%%%%%%%%%%%%%%%%%%%%%%%%%%%%%%%%%%%%%%%%%%%%%%%%%%%%%%
\section{Teleparallel gravity}\label{sec:TG}
Teleparallel gravity can be viewed as a gauge theory of the translation group \cite{Cho:1975dh,Hayashi:1967se,Hayashi:1977jd,Blagojevic,AP}, which allows us to understand the gravitational interaction in analogy to electrodynamics and Yang-Mills theories. We start with a review of the ordinary teleparallel gravity, highlight its analogy with Maxwell electrodynamics, and briefly discuss some of its aspects in which it is more akin to Yang-Mills theories. Then we proceed with the construction of the analogue of teleparallel gravity to the premetric formulation of electrodynamics. It was introduced recently \cite{Itin:2016nxk} for the class of local and linear, i.e., premetric, gravitational constitutive relations, and we will generalize this formulation and construction to general non-linear constitutive relations.

%+++++++++++++++++++++++++++++++++++++++++++++++++%
\subsection{Teleparallel Equivalent of General Relativity}\label{ssec:TPFieldS}
The fundamental variables in teleparallel gravity are a frame $\{e_\mu\}_{\mu=0}^3$ or, equivalently, a coframe $\{\theta^\mu\}_{\mu=0}^3$, which form local bases of $T_xM$ and $T^*_xM$, and a curvature-free spin connection $\omega^\mu{}_\nu$ with torsion. In a local coordinate basis these frame fields may be expanded as
\begin{align}
	e_\mu = e_\mu{}^a \partial_a, \qquad \theta^\mu = \theta^\mu{}_a \dd x^a\,,
\end{align}
and their components are defined to be inverse and so can be related to each other with help of the identities
\begin{equation}\label{key}
\theta^\mu{}_a\, e_\nu{}^a=\delta^\mu_{\nu}, \qquad
\theta^\mu{}_a\, e_\mu{}^b=\delta^b_{a}.
\end{equation}
The components can be used to convert spacetime indices into Lorentz indices, and vice versa, e.g.
\begin{equation}%\label{key}
X^{a_1\cdots}{}_{b_1\cdots}{}^{\mu_1\cdots}{}_{\nu_1\cdots}=
e_{\rho_1}{}^{a_1}\cdots \theta^{\sigma_1}{}_{b_1}\cdots
\theta^{\mu_1}{}_{c_1}\cdots   e_{\nu_1}{}^{d_1}\cdots
X^{\rho_1\cdots}{}_{\sigma_1\cdots}{}^{c_1\cdots}{}_{d_1\cdots}
\end{equation}
where $X^{\cdots}{}_{\cdots}$ is a general tensor with mixed indices. In addition the coframe defines a Lorentzian spacetime metric
\begin{align}\label{eq:metric}
	g_{ab} = \eta_{\mu\nu}\theta^\mu{}_a\theta^\nu{}_b
\end{align}
which is used to raise and lower spacetime indices. From now on, as it is common in the literature, we will refer to both frames and coframes, as well as their components, as \textit{tetrads}.

The first Cartan structure equation defines the torsion as a covariant exterior derivative of a tetrad,
\begin{align}\label{Dtheta=0}
T^\mu\equiv	\mathrm{D} \theta^\mu = \mathrm{d} \theta^\mu + \omega^\mu{}_\nu \wedge \theta^\nu,
\end{align}
where $T^\mu$ are $2$-forms, which in local coordinates can be expressed as
\begin{align}
T^\mu = \frac{1}{2} T^\mu{}_{ab}\mathrm{d}x^a \wedge \mathrm{d}x^b\,,
\end{align}
while the connection coefficients $\omega^\sigma{}_\nu$ are $1$-forms
\begin{align}
\omega^\sigma{}_\nu = \omega^\sigma{}_{\nu a}  \mathrm{d} x^a\,,
\end{align}
which satisfy the condition of vanishing curvature expressed by the second Cartan structure equation
\begin{equation}\label{zerocurv}
\D\omega^\mu{}_\nu = \mathrm{d} \omega^\mu{}_\nu + \omega^\mu{}_\rho \wedge \omega^\rho{}_\nu \equiv 0.
\end{equation}
Based on analogy with $F=\mathrm{d}A$ we can view the torsion as an analogue of the electromagnetic field strength and the tetrad as an analogue of the electromagnetic potential.
Taking the exterior covariant derivative of the torsion \eqref{Dtheta=0} and using \eqref{zerocurv} we find the Bianchi identity
\begin{align}\label{eq:DT=0}
\D T^\mu = 0
\end{align}
that plays the role of the first Maxwell equation in \eqref{eq:dF=0dH=J}.

To derive the field equations of teleparallel gravity we introduce the teleparallel gravity excitation $2$-forms $\tilde{H}_\mu$ as function of the tetrad and the torsion given by
\begin{equation}\label{CRTEGR}
\tilde{H}_\mu(\theta^\nu, T^\nu) =\frac{1}{4} |\theta| \epsilon_{abcd} S_\mu{}^{cd} \dd x^a \wedge \dd x^b\,,
\end{equation}
where $|\theta| \equiv |\det \theta^\mu(\partial_a)| = |\det \theta^\mu{}_a|$ and we expressed its components here in terms of the more commonly used superpotential
\begin{equation}\label{eq:super}
	S_\mu^{\ ab}=\frac{1}{2} \left( T^{ba}{}_{\mu} +T_\mu{}^{ab} -T^{ab}{}_{\mu} \right) -e_\mu^{\ b}T^{ca}{}_{c} + e_\mu^{\ a}T^{cb}{}_{c}\,,
\end{equation}
which can be understood in terms of a linear operator $\tilde \chi$ acting on the torsion
\begin{equation}\label{eq:supercondens}
	S_\mu{}^{ab}  = \Bigg( e_{\sigma_1}{}^{[b}g^{a]c_1} e_\mu{}^{d_1} + \frac{1}{2}\eta_{\mu\sigma_1}g^{ac_1}g^{b{d_1}} + 2 e_\mu{}^{[a} g^{b]c_1} e_{\sigma_1}{}^{d_1}\Bigg) T^{\sigma_1}{}_{c_1d_1}
	= \frac{1}{2}\frac{1}{|\theta|} \tilde \chi_\mu{}^{ab}{}_{\nu}{}^{cd}(\theta)T^\nu{}_{cd}\,.
\end{equation}
This representation is the gravitational analogue of the expansion of the electromagnetic excitation in terms of the electromagnetic field strength in Maxwell electrodynamics, see \eqref{eq:edynloclin} and \eqref{eq:maxconlaw}, as well as the discussion in \cite{Itin:2016nxk}. As in the electrodynamics case the dimensionful constant which needs to appear in the relation between superpotential and torsion is set to one by our choice of units. Up to the density factor $|\theta|$ the operator $\tilde \chi$ appeared in the literature in analysis of propagating degrees of freedom~\cite{Ong:2013qja} and the Hamiltonian formulation of teleparallel gravity~\cite{Ferraro:2016wht}.\footnote{Note that in the literature one also encounters the so called Lucas-Pereira dual $\ast$ on soldered bundles~\cite{Lucas:2008gs}. It allows to write the teleparallel excitation form in a very compact way~$\tilde H_\mu= \eta_{\mu\sigma} (\ast T)^\sigma$, which resembles the constitutive relation of Maxwell electrodynamics \eqref{MaxED}. It is a map from tangent space valued $2$-forms to tangent space valued $2$-forms and takes into account the soldered nature of the tangent bundle, i.e., allows contractions between Lorentz and spacetime indices.}

The teleparallel action, including the coupling to matter fields $\Psi_I$ and setting the gravitational constant $\frac{8 \pi G}{c^4} = 1$, can then be written as
\begin{align}\label{eq:TPaction}
\tilde S[\theta^\nu, \omega^{\rho}{}_{\sigma},\Psi_I]  = \tilde S_g[\theta^\nu, \omega^{\rho}{}_{\sigma}] + S_m[\theta^\nu,\Psi_I] \,,
\end{align}
where the gravitational action is defined by the torsion scalar $\mathbb{T}$
\begin{align}\label{eq:TEGRAction}
\tilde S_g
= \frac{1}{2} \int_M  T^\mu \wedge \tilde{H}_\mu
= \frac{1}{4} \int_M |\theta|\ T^\mu{}_{ab} S_\mu{}^{ab}\ \dd^4 x
\equiv \frac{1}{2} \int_M |\theta|\ \mathbb{T}\ \dd^4 x\,.
\end{align}
We remark that it is possible to derive this action and the above postulated superpotential directly from the Einstein--Hilbert action. One may rewrite the latter in terms of the tetrad, instead of in terms of the metric, and directly obtain the action \eqref{eq:TEGRAction}. The canonical momentum of the tetrad $\theta^\mu$ is then given by the teleparallel gravity excitation $2$-forms $\tilde{H}_\mu$ and the superpotential \eqref{eq:super} is its Hodge dual.

The matter action which describes the coupling between gravity and all other physical fields is given by
\begin{align}
	S_m = \int_M \mathcal{L}_m(\theta^\nu, \Psi_I) \dd^4x\,,
\end{align}
Observe that the matter action we consider only depends on the tetrad and the matter fields and not on the teleparallel spin connection, i.e., we still consider matter fields which are minimally coupled to the metric only. Note that this assumption does not exclude spinning matter, but simply means that spinning matter couples via the Levi-Civita connection derived from the metric, in the same way as in general relativity.

Varying the total action \eqref{eq:TPaction} with respect to the tetrad we find the field equations
\begin{equation}\label{TEGRfe}
\D \tilde{H}_{\mu} - \tilde \Upsilon_{\mu} = \Sigma_{\mu},
\end{equation}
where
\begin{align}
\tilde \Upsilon_{\sigma} &= \frac{1}{48}\epsilon^{abcd}T^{\mu}{}_{cd}\frac{\delta \tilde H_{\mu ab}}{\delta \theta^\sigma{}_r}\epsilon_{rklm}\ \dd x^k \wedge \dd x^l \wedge \dd x^m\,,\label{eq:Upsilon}\\
\Sigma_\mu &= \frac{1}{6} \frac{\partial \mathcal{L}_m}{\partial \theta^\mu{}_a}\epsilon_{abcd} \dd x^b \wedge \dd x^c \wedge \dd x^d\label{EMtensordef}
\end{align}
are known as the gravitational and matter energy-momentum 3-forms, respectively. Note that local Lorentz invariance of the matter action implies that the corresponding energy-momentum tensor is symmetric~\cite{AP,Obukhov:2006ge},
\begin{equation}\label{eq:TPemconserv}
\Sigma^{[\mu}\wedge\theta^{\nu]}=0\,.
\end{equation}

The variation with respect to the spin connection vanishes identically \cite{Krssak:2015lba}. Therefore, the spin connection does not have dynamics on its own, but it plays an important role in keeping the theory locally Lorentz invariant and defining correct conserved charges and the finite action \cite{Lucas:2009nq,Krssak:2015rqa,Krssak:2015lba}.

Having all the fundamental defining relations of teleparallel gravity at hand we can observe the following analogy with Maxwell electrodynamics: Bianchi identities \eqref{eq:DT=0} look like the first equation in \eqref{eq:dF=0dH=J}, the field equations \eqref{TEGRfe} like the second equation in \eqref{eq:dF=0dH=J}, and the constitutive relations \eqref{CRTEGR} and \eqref{MaxED} also have a similar form. However, one can also notice two differences:

Firstly, in teleparallel gravity appears the exterior covariant derivative while in electromagnetism the exterior derivative is the ordinary one\footnote{In \cite{Itin:2016nxk}, authors considered a special gauge $\omega^{\mu}{}_{\nu}=0$, in which the covariant exterior derivative reduces to the ordinary one. As we argue later, it is not necessary to consider this gauge and we can work in a fully covariant theory.}. This is due to the fact that the tetrad, and hence also the torsion, carries a representation of an external gauge group, given by the Lorentz group and gauged with the spin connection \(\omega^{\mu}{}_{\nu}\). This is not the case for the electromagnetic vector potential and field strength.

Secondly, there appears the gravitational energy-momentum $\tilde \Upsilon_\mu$ on the left hand side of the equation \eqref{TEGRfe}. This term appears here as a consequence of the constitutive relation \eqref{CRTEGR} depending on the dynamical field, i.e., the tetrad, itself. From a physical viewpoint it represents self-interaction and is related to the non-linear nature of the gravitational interaction in general relativity\footnote{Note that one should distinguish between the linearity of the interaction and constitutive relation, which are two different concepts. For example, the teleparallel equivalent of general relativity is defined by the constitutive relation \eqref{CRTEGR}, which is linear in the torsion, but the gravitational interaction is determined by non-linear equations of motion, since the tetrad itself enters the constitutive relation as well.}.
In principle, one can consider the total energy-momentum $\Sigma^{(\textrm{total})}_{\mu}$ that includes both the gravitational and matter contributions, and rewrite the field equations \eqref{TEGRfe} as
\begin{align}\label{eq:GravFieldEqHomog}
	\D \tilde{H}_{\mu} = \tilde \Upsilon_\mu + \Sigma_\mu = \Sigma^{(\textrm{total})}_{\mu}.
\end{align}
In this form the similarity to the corresponding electromagnetic field equation $\mathrm{d} H=J$ becomes the most apparent.

The presence of both the covariant exterior derivatives and the self-interaction terms imply that the teleparallel gravity is actually more similar to the Yang-Mills gauge theory than the Maxwell electrodynamics. Nevertheless, the analogy with electrodynamics is still sufficient for our main task here: to construct generalized gravity models by different choices of constitutive laws as it is done in premetric electrodynamics.

%+++++++++++++++++++++++++++++++++++++++++++++++++%
\subsection{Teleparallel gravity with general constitutive relation}\label{ssec:TPInd}
The teleparallel equivalent of GR, ordinary teleparallel gravity, which we discussed in the previous section, was recently generalized in a similar fashion as premetric electrodynamics generalizes Maxwell electrodynamics \cite{Itin:2016nxk}. These local and linear teleparallel gravity theories, i.e., the gravitational excitation depends linearly on the torsion tensor, cover basically what is known as new GR theories in the literature \cite{Hayashi:1979qx}.

In order to extend the use of the language of premetric electrodynamics from local and linear theories covered in \cite{Itin:2016nxk} to local, but in general non-linear teleparallel gravity theories, we now consider gravitational excitation tensors $H_\mu$ which are general functions of the tetrad (the potential of the theory) and the torsion (the field strength of the potential),
\begin{align}\label{CRPATG}
H_\mu = H_\mu(\theta^\nu, T^\nu)\,.
\end{align}
Observe that there is an important difference between our formulation of gravity in terms of constitutive laws and electrodynamics. In contrast to electrodynamics the gravitational excitation depends not only on the corresponding field strength, but also on the potential itself. The feature already appeared for the teleparallel equivalent of GR in the previous section and is the source of the gravitational energy term $\tilde \Upsilon_\mu$, which will appear in the field equations in general for gravity, but is absent in electrodynamics.

The Bianchi identities \eqref{eq:DT=0} and the matter energy-momentum \eqref{EMtensordef} with the conservation laws \eqref{eq:TPemconserv} do not involve the excitation form and hence stay the same as in the previous section. Only the field equations \eqref{TEGRfe} change due to the generalization of the constitutive relation \eqref{CRPATG}.

The details of the calculation can be found in Appendix~\ref{app:feqder}, where we consider the generalized gravitational action
\begin{equation}\label{GravAc}
S_g[\theta^\nu, \omega^{\rho}{}_{\sigma}] = \frac{1}{2} \int_M T^\mu \wedge {H}_\mu
\end{equation}
in \eqref{eq:TPaction} instead of $\tilde S_g$. Variation of the action $S=S_g+S_m$ yields the new extended field equations
\begin{align}\label{eq:GravFieldEquations1}
\D\Pi_{\mu} - \Upsilon_{\mu} = \Sigma_{\mu}\,,
\end{align}
where we introduced the $2$-form
\begin{align}
	\Pi_\sigma &= \frac{1}{2}(H_\sigma + Q_\sigma) = \frac{1}{2}H_\sigma + \frac{1}{16} \epsilon^{abcd}T^\mu{}_{ab}\frac{\delta H_{\mu cd}}{\delta T^\sigma{}_{rs}} \epsilon_{rskl}\ \mathrm{d}x^k \wedge \mathrm{d}x^l \label{eq:Pi},
\end{align}
which is the canonical momentum of the tetrad, as can be seen from the form of the variation of the action in \eqref{eq:deltaSG}. Note that in the TEGR case, discussed in the previous section, $\Pi_\sigma = H_\sigma = \tilde H_\sigma$, while in general $\Pi_\sigma$ is different from $H_\sigma$ due its possible non-linear dependence on the torsion. The $3$-form $\Upsilon_\mu$ is defined analogous to $\tilde \Upsilon_\mu$, see \eqref{eq:Upsilon}, by replacing $\tilde H_{\mu ab}$ with $H_{\mu ab}$,
\begin{align}
	 \Upsilon_{\sigma} &= \frac{1}{48}\epsilon^{abcd}T^{\mu}{}_{cd}\frac{\delta H_{\mu ab}}{\delta \theta^\sigma{}_r}\epsilon_{rklm}\ \dd x^k \wedge \dd x^l \wedge \dd x^m \label{eq:UpsilonNew}\,.
\end{align}
Note that for explicit calculations the variational derivative is given by
\begin{equation}
\frac{\delta T^\mu{}_{ab}}{\delta T^\sigma{}_{cd}} = \delta^{\mu}_\sigma \delta^c_{[a}\delta^d_{b]}\,,
\end{equation}
which is different from a simple derivative with respect to the components of the torsion, and takes into account the antisymmetry in the last two indices.

The presence of the extra term $Q_\sigma$ makes the form of the field equations differ slightly from the ones in ordinary teleparallel gravity \eqref{TEGRfe}. However, for excitation $2$-forms $H_\mu(\theta^\nu, T^\nu)$ which are homogeneous of any degree $r$ in the torsion, i.e., if $H_\mu(\theta^\nu,kT^\nu) = k^r H_\mu(\theta^\nu, T^\nu)$ holds for \(k \in \mathbb{R}\), it turns out that $Q_\mu = r H_\mu$. Hence, for these models the field equations take the same form as in the case of a linear gravitational constitutive relation, up to constant factor.

Observe that the same is true when one derives the field equations of electrodynamics from an action
\begin{equation}
\int_M \left[\frac{1}{2} F \wedge H(F) + A\wedge J\right]\,.
\end{equation}
The equations of motion only have the form $\dd H = J$ if $H$ is local and linear in $F$, otherwise one finds $d\Pi = J$ with $\Pi =\frac{1}{2}( H + Q)$, where $Q$ contains derivatives of $H$ with respect to $F$.

The field equations \eqref{eq:GravFieldEquations1} can be decomposed into their symmetric part
\begin{equation}
\D\Pi^{(\mu} \wedge \theta^{\nu)} - \Upsilon^{(\mu} \wedge \theta^{\nu)} = \Sigma^{(\mu}\wedge\theta^{\nu)},\label{eq:GravFieldEquations2}
\end{equation}
and their antisymmetric part
\begin{equation}
\D\Pi^{[\mu} \wedge \theta^{\nu]} - \Upsilon^{[\mu} \wedge \theta^{\nu]} = 0\,,\label{eq:GravFieldEquations3}
\end{equation}
where we have used the symmetricity of the matter energy-momentum tensor
\eqref{eq:TPemconserv}.

Differently to the case of the ordinary teleparallel gravity, the variation with respect to the spin connection does not vanish identically but--as we explicitly show in Appendix~\ref{app:feqder}--it yields the antisymmetric part of the field equations for the tetrad \eqref{eq:GravFieldEquations3}. This antisymmetric part can then equally be viewed as the ``field equations'' for the spin connection. However, since it does not contain second derivatives of the spin connection, it rather yields constraints than equations which encode new dynamics.

For spin connections which solve the constraint \eqref{eq:GravFieldEquations3} only the symmetric part of the field equations \eqref{eq:GravFieldEquations2} need to be solved. A possible method to obtain spin connections which satisfy the constraint explicitly was recently discussed for $f(\mathbb{T})$ gravity in \cite{Krssak:2015oua,Krssak:2017nlv}.

%%%%%%%%%%%%%%%%%%%%%%%%%%%%%%%%%%%%%%%%%%%%%%%%%%%%%%%
\section{Gravitational constitutive relations}\label{sec:GravConst}
The advantage of the generalization of teleparallel gravity in the language of general constitutive laws is that it allows us to study properties of various extended models of gravity in a unified way. The field equations of these models always take the form
\begin{align}\label{FirstEq}
\D T^\mu &= 0\,,\\
\D\Pi_{\mu} - \Upsilon_{\mu}& = \Sigma_{\mu}\label{SecondEq}
\end{align}
and differ only in the constitutive relation defining the particular model. The different building blocks of the equations were defined in the previous section, see \eqref{eq:Pi} for the highest derivative term $\Pi_\mu$, \eqref{eq:Upsilon} for the gravitational self energy $\Upsilon_\mu$ and \eqref{EMtensordef} for the matter energy momentum $\Sigma_\mu$.

Here we now present different teleparallel gravity constitutive relations and demonstrate how they realize various popular modified teleparallel gravity models such as $f(\mathbb{T})$ gravity in this framework.

%+++++++++++++++++++++++++++++++++++++++++++++++++%
\subsection{Polynomial teleparallel constitutive relations}
A very convenient choice of constitutive relation is a polynomial in the torsion tensors which are contracted by a constitutive tensor $\kappa$. This constitutive tensor can be built from the tetrads $\theta^\nu$. The components of $H_\mu$ take the form
\begin{align}
	H_{\mu ab}(\theta^\nu, T^\nu)
	&= \frac{1}{2^r}\kappa_{\mu ab}{}_{\sigma_1}{}^{c_1d_1} {}_{\sigma_2}{}^{c_2d_2} {}_{....}{}^{...}{}_{\sigma_r}{}^{c_rd_r}(x,\theta^\nu)T^{\sigma_1}{}_{c_1d_1}T^{\sigma_2}{}_{c_2d_2}...T^{\sigma_r}{}_{c_rd_r}\nonumber\\
	&= \frac{1}{2^{r+1}}\epsilon_{abcd}\chi_{\mu}{}^{cd}{}_{\sigma_1}{}^{c_1d_1} {}_{\sigma_2}{}^{c_2d_2} {}_{....}{}^{...}{}_{\sigma_r}{}^{c_rd_r}(x,\theta^\nu)T^{\sigma_1}{}_{c_1d_1}T^{\sigma_2}{}_{c_2d_2}...T^{\sigma_r}{}_{c_rd_r}\,,
\end{align}
where we introduced the constitutive density $\chi_{\mu}{}^{ab}{}_{\sigma_1}{}^{c_1d_1} {}_{\sigma_2}{}^{c_2d_2} {}_{....}{}^{...}{}_{\sigma_r}{}^{c_rd_r}(x,\theta^\nu)$. From the form of the action it is clear that the constitutive density $\chi$ has a pairwise triple exchange symmetry, i.e., is symmetric under the pairwise exchange of any triple of indices ${}_{\sigma_i}{}^{p_i q_i}$ including the free indices ${}_{\mu}{}^{a b}$
\begin{align}
	\chi_{\mu}{}^{cd}{}_{\sigma_1}{}^{c_1d_1} {}_{\sigma_2}{}^{c_2d_2} {}_{....}{}^{...}{}_{\sigma_r}{}^{c_rd_r}&= \chi_{\mu}{}^{cd}{}_{\sigma_2}{}^{c_2d_2} {}_{\sigma_1}{}^{c_1d_1} {}_{....}{}^{...}{}_{\sigma_r}{}^{c_rd_r}= ... \nonumber\\
	= \chi_{\sigma_1}{}^{c_1d_1} {}_{\mu}{}^{cd} {}_{\sigma_2}{}^{c_2d_2} {}_{....}{}^{...}{}_{\sigma_r}{}^{c_rd_r} &= \chi_{\mu}{}^{cd}{}_{\sigma_1}{}^{c_1d_1}{}_{\sigma_r}{}^{c_rd_r} {}_{....}{}^{...} {}_{\sigma_2}{}^{c_2d_2}\,.
\end{align}
The $2$-forms $Q_\sigma$ can now be calculated and become
\begin{align}
Q_\sigma = \frac{1}{8} \epsilon^{abcd} T^{\mu}{}_{ab}\frac{\delta H_{\mu cd}}{\delta T^\sigma{}_{rs}} \epsilon_{rskl}\ \dd x^k \wedge \dd x^l =r H_\sigma\,,
\end{align}
so that the field equations are given by
\begin{align}
	\D H_\mu =  \frac{2}{(1+r)}(\Upsilon_\mu + \Sigma_\mu)\,,
\end{align}
which we recognize to be a special instance of the field equations for general homogeneous constitutive relations.
The gravitational self interaction energy $3$-forms $\Upsilon_\mu$ depend on how the constitutive tensor $\kappa$ is constructed from the tetrad.

For $r=1$ one obtains the class of constitutive relations which were discussed in \cite{Itin:2016nxk} yielding the new GR models~\cite{Hayashi:1979qx},
\begin{align}
	H_{\mu ab}(\theta^\nu, T^\nu)
	= \frac{1}{2}\kappa_{\mu ab}{}_{\sigma_1}{}^{c_1d_1}(x,\theta^\nu) T^{\sigma_1}{}_{c_1d_1}
	= \frac{1}{4}\epsilon_{abcd}\chi_{\mu}{}^{cd}{}_{\sigma_1}{}^{c_1d_1}(x,\theta^\nu) T^{\sigma_1}{}_{c_1d_1}\,.
\end{align}
Comparing this expression with the induction form of the teleparallel equivalent of GR \eqref{CRTEGR}, we can read off the corresponding constitutive density in terms of the superpotential, which itself is linear in the torsion, or as contraction operator built from the tetrads acting on the torsion \eqref{eq:supercondens}, where the metric is understood as function of the tetrads, see~\eqref{eq:metric}
\begin{align}\label{eq:TEGRconst}
	\frac{1}{2}\chi_{\mu}{}^{cd}{}_{\sigma_1}{}^{c_1d_1}(\theta^\nu) T^{\sigma_1}{}_{c_1d_1}
	= |\theta|\Bigg( e_{\sigma_1}{}^{[d}g^{c]c_1} e_\mu{}^{d_1} + \frac{1}{2}\eta_{\mu\sigma_1}g^{cc_1}g^{d{d_1}} + 2 e_\mu{}^{[c} g^{d]c_1} e_{\sigma_1}{}^{d_1}\Bigg) T^{\sigma_1}{}_{c_1d_1} = |\theta| S_\mu{}^{cd}\,.
\end{align}

%+++++++++++++++++++++++++++++++++++++++++++++++++%
\subsection{$f(\mathbb{T})$ gravity}
The most studied extension of teleparallel gravity is the so-called $f(\mathbb{T})$ gravity \cite{Ferraro:2006jd,Ferraro:2008ey,Bengochea:2008gz,Linder:2010py} constructed from an arbitrary function of the torsion scalar
\begin{equation}\label{TGscalar}
\mathbb{T}= \frac{1}{2} S_\mu{}^{ab} \, T^\mu{}_{ab}\,.
\end{equation}
It can be read off from the action \eqref{eq:TEGRAction} and the superpotential $S_\mu{}^{ab}$ was displayed in \eqref{eq:super}.

One easily checks that $f(\mathbb{T})$ gravity can be realized in our approach using a constitutive relation of the form
\begin{equation}\label{key1}
H_{\mu ab}(\theta^\nu, T^\nu)= \frac{1}{2} \frac{f(\mathbb{T})}{\mathbb{T}} |\theta| \epsilon_{abcd} S_\mu{}^{cd}\,.
\end{equation}
Note that one of the original motivations to consider $f(\mathbb{T})$ gravity \cite{Ferraro:2008ey} was the construction of a gravitational analogue of Born-Infeld theory by taking the function
\begin{equation}\label{key2}
f(\mathbb{T})=\epsilon  \left(\sqrt{1+\frac{2 \mathbb{T}}{\epsilon}}\right),
\end{equation}
where $\epsilon$ is a constant controlling the scale of the deformation. Within our framework inspired by non-linear electrodynamics such a construction appears naturally and can be easily understood.

%+++++++++++++++++++++++++++++++++++++++++++++++++%
\subsection{Gravitational analogue of Pleba\'nski electrodynamics}
Recently \cite{Bahamonde:2017wwk}, an even larger class of modified teleparallel gravity theories--so-called $f(T_\text{ax},T_\text{ten},T_\text{vec})$ gravity--were proposed, which include $f(\mathbb{T})$-gravity and other relevant models in the literature, such as teleparallel conformal gravity~\cite{Maluf:2011kf}. We now show how $f(T_\text{ax},T_\text{ten},T_\text{vec})$ gravity naturally fits in our new framework and can be realized as special case of the gravitational analogue of Pleba\'nski electrodynamics, discussed in section~\ref{SecED}, which we develop here.

To construct this analogue, we consider all quadratic invariants of torsion. It turns that it is more suitable to first decompose the torsion tensor into irreducible pieces with respect to the Lorentz group as
\begin{align}
T_{abc} = \frac{2}{3}(t_{abc}-t_{acb}) +
\frac{1}{3}(\eta_{ab}v_{c}-\eta_{ac}v_{b}) +
\epsilon_{abcd}a^{d}\,,
\end{align}
where
\begin{align}
v_{a} = T^{b}{}_{ba}\qquad
a_{b} = \frac{1}{6}\epsilon_{bcde}T^{cde}\,,\qquad
t_{abc} = \frac{1}{2}(T_{abc}+T_{bac}) +
\frac{1}{6}(g_{ca}v_{b}+g_{cb}v_{a})-\frac{1}{3}g_{ab}v_{c}\,,
\end{align}
are known as the vector, axial, and purely tensorial torsions, respectively.

The advantage of this decomposition is that we can then distinguish between parity preserving invariants that we denote as
\begin{align}\label{inv13}
T_{\rm ax} = a_{b}a^{b}\,, \qquad
T_{\rm ten} = t_{abc}t^{abc}\,, \qquad
T_{\rm vec} = v_{b}v^{b}\,,
\end{align}
and parity violating invariants
\begin{equation}\label{inv45}
I_4=v^b a_b \,, \qquad \text{and} \qquad
I_5=\epsilon_{abcd} t^{e ab} t_e{}^{cd} \,.
\end{equation}
The gravitational analogue of Pleba\'nski electrodynamics \eqref{CRPleba\'nski} is now obtained from the gravitational constitutive relation
\begin{align}\label{CRFTTT}
	H_{\mu ab}{}(\theta^\nu, T^\nu) =\frac{1}{2} U(T_{\rm ax}, T_{\rm ten}, T_{\rm vec}, I_4, I_5)\, T_{\mu ab}  + \frac{1}{2} V(T_{\rm ax}, T_{\rm ten}, T_{\rm vec}, I_4, I_5)\, |\theta|\epsilon_{abcd} S_\mu{}^{cd}\,,
\end{align}
where $U$ and $V$ are arbitrary functions of five invariants \eqref{inv13}-\eqref{inv45}.

It becomes clear now that $f(T_\text{ax},T_\text{ten},T_\text{vec})$ gravity can be realized by restricting to the parity preserving invariants \eqref{inv13} and choosing functions $U$ and $V$ as
\begin{equation}
U=0,\qquad V=\frac{f(T_\text{ax},T_\text{ten},T_\text{vec})}{\mathbb{T}}.
\end{equation}

With the constitutive relation \eqref{CRFTTT} we further enlarged the class of possibly interesting teleparallel gravity models which may be investigated for their physical relevance in the future. The striking insight of our construction is that it is possible to analyse all of these different theories of gravity by a set of field equations which have a common universal form \eqref{eq:GravFieldEquations1}.

%+++++++++++++++++++++++++++++++++++++++++++++++++%
\subsection{Teleparallel dark energy}
Another popular model is the so-called teleparallel dark energy model \cite{Geng:2011aj}, where the torsion scalar \eqref{TGscalar} is non-minimally coupled to a scalar field. This model can be realized through an extension of our framework in terms of an excitation tensor which further depends on a scalar field $\phi$ and a constant parameter $\xi$,
\begin{equation}\label{key3}
H_{\mu ab}(\theta^\nu, T^\nu, \phi)= \frac{1}{4}|\theta| \epsilon_{abcd} (1+\phi^2 \xi) S_\mu{}^{cd}\,.
\end{equation}
The dynamics of the scalar field is given by an additional field equation. It is obtained by including a kinetic term as well as a potential term for the scalar field to the action \eqref{GravAc} and performing the corresponding variation.

The teleparallel dark energy constitutive density then also contains the scalar field,
\begin{align}
\frac{1}{2}\chi_{\mu}{}^{cd}{}_{\sigma_1}{}^{c_1d_1}(\theta^\nu,\phi) = |\theta|(1 + \phi^2\xi) \Bigg( e_{\sigma_1}{}^{[d}g^{c]c_1} e_\mu{}^{d_1} + \frac{1}{2}\eta_{\mu\sigma_1}g^{cc_1}g^{d{d_1}} + 2 e_\mu{}^{[c} g^{d]c_1} e_{\sigma_1}{}^{d_1}\Bigg)\,.
\end{align}
In electrodynamics, theories with different fields in the constitutive relation describe for example the uniaxial crystal, see \eqref{eq:uniax}.

%%%%%%%%%%%%%%%%%%%%%%%%%%%%%%%%%%%%%%%%%%%%%%%%%%%%%%%
\section{Discussion}\label{sec:disc}
We have developed a general framework in which teleparallel theories of gravity can be formulated in analogy to various theories of electrodynamics. Our building blocks are: the torsion tensor, which is the gravitational analogue of the field strength, an excitation tensor, and a gravitational constitutive relation expressing the excitation as a function of the torsion and the tetrad.
The gravitational dynamics then are described by the Bianchi identities \eqref{eq:DT=0} and the field equations \eqref{eq:GravFieldEquations1}, which we have derived for a general constitutive relation.

We realized various teleparallel models like $f(\mathbb{T})$ gravity, $f(T_\text{ax},T_\text{ten},T_\text{vec})$ gravity and teleparallel dark energy model within our framework by explicitly stating their constitutive relation. Moreover we outlined how to construct gravity theories in analogy to theories of electrodynamics and demonstrated this procedure by proposing a new class of gravity theories based on analogy with Pleba\'nski electrodynamics \eqref{CRFTTT}.

One central advantage of our framework is that the gravitational field equations have always the same compact form and different models of gravity are defined by the choice of a constitutive relation. This allows us to systematically study modified teleparallel theories. A particularly important systematic result we derived here is that the variation of the action with respect to the spin connection yields a constraint equation that coincides with the antisymmetric part of the tetrad field equations \eqref{eq:GravFieldEquations3}. To obtain a consistent solution of the theory the spin connection and the tetrad must be chosen such that this constraint is satisfied, i.e., the spin connection cannot be chosen arbitrarily. These findings agree with the result of \cite{Golovnev:2017dox,Krssak:2017nlv} for $f(\mathbb{T})$ gravity, but now can be understood as a generic feature of all teleparallel theories with second order field equations.

Further our framework allows for a new systematic classification of modified gravity theories according to their defining constitutive relation. The task now is to systematically identify those constitutive relations which lead to theories of gravity with correct (post) Newtonian limit, being ghost-free and consistent with observations in general.

A possible extension of our framework is to consider more general constitutive relations which include higher derivatives of the tetrad, respectively derivatives of the torsion. In electrodynamics such constitutive laws are known and appear for example in Bopp-Podolsky electrodynamics; a higher derivative theory of electrodynamics with finite self interaction of the electron with its own electromagnetic field \cite{Bopp,Podolsky,Gratus:2015bea}.

Another direction of generalization is to allow additional fields in constitutive relations. A most simple example was already presented here by using a scalar field in the example of a teleparallel dark energy model \eqref{key3}. However, it can be generalized to include additional background fields that could be used for an effective description of gravitational effects at certain scales and environments. For example, an additional $1$-form may be considered which is present in some phenomenological models of quantum gravity \cite{Barcaroli:2017gvg}.

Going even further, we may address quantum gravity itself in the teleparallel formulation and search for gravitational constitutive laws which could yield new renormalizable theories of gravity. In \cite{Pfeifer:2016har} the quantization of electrodynamics with general local and linear constitutive law was performed. The reformulation of gravity theories in a similar language may have the potential to approach the search for a consistent theory of quantum gravity from a new point of view.

%%%%%%%%%%%%%%%%%%%%%%%%%%%%%%%%%%%%%%%%%%%%%%%%%%%%%%%
\begin{acknowledgments}
We thank F. W. Hehl, O. Vilson and M. J. Guzm\'an for their helpful remarks on our manuscript. The authors were supported by the Estonian Ministry for Education and Science Institutional Research Support Project IUT02-27 and Startup Research Grant PUT790, as well as by the European Regional Development Fund through the Center of Excellence TK133.
\end{acknowledgments}

\appendix

%%%%%%%%%%%%%%%%%%%%%%%%%%%%%%%%%%%%%%%%%%%%%%%%%%%%%%%
\section{Derivation of the field equations}\label{app:feqder}
Here we present the details of the derivation of the teleparallel field equations \eqref{eq:GravFieldEquations1}, \eqref{eq:GravFieldEquations2} and \eqref{eq:GravFieldEquations3} in differential form language.
We assume that the gravitational action is of the form \eqref{GravAc} and hence its variation is given by
\begin{align}\label{var1}
\delta S_g = \frac{1}{2} \int_M (\delta T^\mu \wedge H_{\mu} + T^\mu \wedge \delta H_{\mu}).
\end{align}
Using $H_{\mu}=H_{\mu}(\theta^\nu, T^\nu)$ we can write the second term as
\begin{align}
\frac{1}{2}T^\mu \wedge \delta H_\mu
&= \frac{1}{8}\bigg( T^{\mu}{}_{ab}\epsilon^{abcd}\frac{\delta H_{\mu cd}}{\delta T^\sigma{}_{rs}}\delta T^\sigma{}_{rs} + T^{\mu}{}_{ab}\epsilon^{abcd}\frac{\delta H_{\mu cd}}{\delta \theta^\sigma{}_{r}}\delta \theta^\sigma{}_{r}\bigg) \mathrm{d}x^0 \wedge \mathrm{d}x^1 \wedge \mathrm{d}x^2 \wedge \mathrm{d}x^3\nonumber\\
&= \frac{1}{8}\bigg( \frac{1}{4} T^{\mu}{}_{ab}\epsilon^{abcd}\frac{\delta H_{\mu cd}}{\delta T^\sigma{}_{rs}} \epsilon_{klrs} \epsilon^{klij} \delta T^\sigma{}_{ij}+ \frac{1}{6}  T^{\mu}{}_{ab}\epsilon^{abcd}\frac{\delta H_{\mu cd}}{\delta \theta^\sigma{}_{r}} \epsilon_{klmr} \epsilon^{klmi} \delta \theta^\sigma{}_{i}\bigg)\mathrm{d}x^0 \wedge \mathrm{d}x^1 \wedge \mathrm{d}x^2 \wedge \mathrm{d}x^3\\
& =\frac{1}{2} Q_\sigma \wedge \delta T^\sigma + \Upsilon_\sigma \wedge \delta \theta^\sigma \,,\nonumber
\end{align}
and write \eqref{var1} then as
\begin{align}\label{eq:deltaSG}
\delta S_g = \int_M(\Upsilon_{\mu} \wedge \delta\theta^{\mu} + \Pi_{\mu} \wedge \delta T^{\mu}),
\end{align}
where the (twisted) 3-forms \(\Upsilon_{\mu}\) and (twisted) 2-forms \(\Pi_{\mu}\) are defined in equations \eqref{eq:Pi} and \eqref{eq:UpsilonNew}. Note that $\delta T^\mu$ contains the variation with respect to $\mathrm{d}\theta^\mu$, see equation \eqref{eq:deltaT}. Hence we can identify $\Pi_\mu$ with the canonical momentum associated to $\theta^\mu$.

We further demand that the action is invariant under local infinitesimal Lorentz transformations \(\lambda^{\mu}{}_{\nu}\), where $\lambda^{(\mu\nu)}=0$, which induce the variations \(\delta_{\lambda}\theta^{\mu} = \lambda^{\mu}{}_{\nu}\theta^{\nu}\) and \(\delta_{\lambda}T^{\mu} = \lambda^{\mu}{}_{\nu}T^{\nu}\). One finds that the corresponding variation of the action is given by

\begin{equation}
\delta_{\lambda}S_g = \int_M\left[\Upsilon_{\mu} \wedge (\lambda^{\mu}{}_{\nu}\theta^{\nu}) + \Pi_{\mu} \wedge (\lambda^{\mu}{}_{\nu}T^{\nu})\right] = \int_M\left(\Upsilon^{[\mu} \wedge \theta^{\nu]} + \Pi^{[\mu} \wedge T^{\nu]}\right)\lambda_{\mu\nu}\,.
\end{equation}

It follows that the action is locally Lorentz invariant if and only if
\begin{equation}\label{eqn:lorinvgrav}
\Upsilon^{[\mu} \wedge \theta^{\nu]} + \Pi^{[\mu} \wedge T^{\nu]} = 0\,.
\end{equation}
Note that this must hold also off-shell. To derive the field equations, we decompose the variation of the torsion as
\begin{equation}\label{eq:deltaT}
\delta T^{\mu} = \delta \D\theta^{\mu} = \mathrm{d}\delta\theta^{\mu} + \omega^{\mu}{}_{\nu} \wedge \delta\theta^{\nu} + \delta\omega^{\mu}{}_{\nu} \wedge \theta^{\nu}\,
\end{equation}
and find that the variation of the total action $S=S_g + S_m$, including also the matter part, is given by
\begin{equation}
\begin{split}
\delta S &= \int_M\left[\Upsilon_{\mu} \wedge \delta\theta^{\mu} + \Pi_{\mu} \wedge (\mathrm{d}\delta\theta^{\mu} + \omega^{\mu}{}_{\nu} \wedge \delta\theta^{\nu} + \delta\omega^{\mu}{}_{\nu} \wedge \theta^{\nu}) + \Sigma_{\mu} \wedge \delta\theta^{\mu}\right]\\
&= \int_M\left[(\Upsilon_{\mu} - \mathrm{d}\Pi_{\mu} + \omega^{\nu}{}_{\mu} \wedge \Pi_{\nu} + \Sigma_{\mu}) \wedge \delta\theta^{\mu} - \Pi_{\mu} \wedge \theta^{\nu} \wedge \delta\omega^{\mu}{}_{\nu}\right]\\
&= \int_M\left[(\Upsilon_{\mu} - \D\Pi_{\mu} + \Sigma_{\mu}) \wedge \delta\theta^{\mu} - \Pi_{\mu} \wedge \theta^{\nu} \wedge \delta\omega^{\mu}{}_{\nu}\right]\,.
\end{split}
\end{equation}
From the variation with respect to the tetrad we now obtain the field equation
\begin{equation}\label{eqn:feqtetrad}
\D\Pi_{\mu} - \Upsilon_{\mu} = \Sigma_{\mu}\,.
\end{equation}
Note that the right hand side, which is given by the energy-momentum 3-form, is symmetric, see equation \eqref{eq:TPemconserv}. To see that also the left hand side is symmetric, we consider the variation \(\delta\omega^{\mu}{}_{\nu}\) of the spin connection. Here we allow only variations which preserve the flatness of the spin connection, and hence satisfy
\begin{equation}
0 = \delta R^{\mu}{}_{\nu} = \mathrm{d}\delta\omega^{\mu}{}_{\nu} + \delta\omega^{\mu}{}_{\rho} \wedge \omega^{\rho}{}_{\nu} + \omega^{\mu}{}_{\rho} \wedge \delta\omega^{\rho}{}_{\nu} = \D\delta\omega^{\mu}{}_{\nu}\,.
\end{equation}
Since the connection is flat, and hence \(\D^2 = 0\), this is solved by \(\delta\omega^{\mu}{}_{\nu} = \D\xi^{\mu}{}_{\nu}\), where \(\xi^{(\mu\nu)} = 0\) in order to preserve the antisymmetry of \(\omega^{\mu}{}_{\nu}\). The corresponding variation of the action is given by
\begin{equation}
\begin{split}
\delta_{\omega}S_G &= -\int_M\Pi_{\mu} \wedge \theta^{\nu} \wedge \D\xi^{\mu}{}_{\nu}\\
&= -\int_M\Pi_{\mu} \wedge \theta^{\nu} \wedge \left(\mathrm{d}\xi^{\mu}{}_{\nu} + \omega^{\mu}{}_{\rho} \wedge \xi^{\rho}{}_{\nu} - \omega^{\rho}{}_{\nu} \wedge \xi^{\mu}{}_{\rho}\right)\\
&= -\int_M\left(\mathrm{d}\Pi_{\mu} \wedge \theta^{\nu} + \Pi_{\mu} \wedge \mathrm{d}\theta^{\nu} + \Pi_{\rho} \wedge \theta^{\nu} \wedge \omega^{\rho}{}_{\nu} - \Pi_{\mu} \wedge \theta^{\rho} \wedge \omega^{\nu}{}_{\rho}\right)\xi^{\mu}{}_{\nu}\\
&= -\int_M\left(\D\Pi_{\mu} \wedge \theta^{\nu} + \Pi_{\mu} \wedge \D\theta^{\nu}\right)\xi^{\mu}{}_{\nu}\,.
\end{split}
\end{equation}
Since the matter action does not depend on \(\omega\), we find the field equation
\begin{equation}\label{eqn:feqconn}
\D\Pi^{[\mu} \wedge \theta^{\nu]} + \Pi^{[\mu} \wedge T^{\nu]} = 0\,.
\end{equation}
Together with the Lorentz invariance condition~\eqref{eqn:lorinvgrav} we thus find that
\begin{equation}
\D\Pi^{[\mu} \wedge \theta^{\nu]} - \Upsilon^{[\mu} \wedge \theta^{\nu]} = 0\,.
\end{equation}
This is simply the condition that also the left hand side of the tetrad field equation~\eqref{eqn:feqtetrad} is symmetric. Note that the same result was obtained in the case of $f(\mathbb{T})$ gravity recently~\cite{Golovnev:2017dox}, but our analysis shows that it holds for all teleparallel theories with the action that can be written as \eqref{GravAc}.

%%%%%%%%%%%%%%%%%%%%%%%%%%%%%%%%%%%%%%%%%%%%%%%%%%%%%%%
\bibliographystyle{utphys}
\bibliography{PreTP,TeleBib}

\end{document}